\documentclass[11pt]{article}
\def\nottoobig#1{{\hbox{$\left#1\vcenter to1.111\ht\strutbox{}\right.\n@space$}}}

\newtheorem{theorem}{Theorem}[section]

\newtheorem{lemma}[theorem]{Lemma}

\newtheorem{definition}[theorem]{Definition}

\usepackage{epsfig}
\usepackage{amsmath}
\usepackage{amsfonts}

\newcommand{\nat}{{\bf N}}

\newcommand{\bpp}{{\rm BPP}}

\newcommand{\proof}{{\bf Proof}}

\def\nottoobig#1{{\hbox{$\left#1\vcenter
to1.111\ht\strutbox{}\right.\n@space$}}}

\newcommand{\prob}{{\rm Prob}}
\newcommand{\sigman}{\Sigma^n}
\newcommand{\sigmal}{\Sigma^l}

\newcommand{\ie}{$\mbox{i.e.}$}

\newlength{\filength}
\newsavebox{\gcbox}
\sbox{\gcbox}{\framebox[\filength]{\rule{0ex}{2ex}}}

\newcommand{\qedblob}{\mbox{\rule[-1.5pt]{5pt}{10.5pt}}}
\def\literalqed{{\ \nolinebreak\hfill\mbox{\qedblob\quad}}}

\def\qed{\literalqed}

\newcommand{\singlespacing}{\let\CS=
\@currsize\renewcommand{\baselinestretch}{1}\tiny\CS}
\newcommand{\singlespacingplus}{\let\CS=
\@currsize\renewcommand{\baselinestretch}{1.25}\tiny\CS}
\newcommand{\doublespacing}{\let\CS=
\@currsize\renewcommand{\baselinestretch}{1.75}\tiny\CS}
\newcommand{\draftspacing}{\let\CS=
\@currsize\renewcommand{\baselinestretch}{2.0}\tiny\CS}

\def\zo{\{0,1\}}

\def\AND{\wedge}

\def\mapping{\rightarrow}

\newcommand{\zon}{\zo^n}
\newcommand{\lll}{\lVert}
\newcommand{\rrr}{\rVert}
\newcommand{\myrule}{\shortrule}
\newcommand{\crc}{{\rm CIRC}}
\newcommand{\perm}{{\rm PERM}}

\newcommand{\shortrule}{
\noindent
\rule{5cm}{0.5pt}
}

\newcommand{\tn}{\tilde{n}}

\newcommand{\barz}{\overline{z}}
\newcommand{\bary}{\overline{y}}
\newcommand{\barx}{\overline{x}}

\newcommand{\barR}{\overline{R}}
\newcommand{\barX}{\overline{X}}
\newcommand{\barN}{\overline{N}}
\newcommand{\barn}{\overline{n}}

\renewcommand{\baselinestretch}{1}

\makeatletter
\def\@listI{\leftmargin\leftmargini \parsep 4.5pt plus 1pt minus 1pt\topsep6pt plus 2pt minus 2pt \itemsep  2pt plus 2pt minus 1pt}

\let\@listi\@listI
\@listi
\makeatother

\author{ {Marius Zimand\/}
\thanks{  Department of Computer and Information Sciences, Towson University,
Baltimore, MD. http://triton.towson.edu/\~{ }mzimand}}

\date{}

\title{Simple extractors via constructions of cryptographic pseudo-random generators}

\begin{document}

\maketitle

{\bf Abstract.} Trevisan has shown that constructions of pseudo-random generators from 
hard functions (the Nisan-Wigderson approach) also produce extractors. 
We show that constructions of pseudo-random generators from one-way permutations 
(the Blum-Micali-Yao approach) can be used for building extractors as well. 
Using this new technique we build extractors that do not use designs or polynomial-based error-correcting codes and that are very simple and efficient. For example, one extractor produces each output bit separately
in $O(\log^2 n)$ time.
These extractors work for weak sources with min entropy $\lambda n$, for arbitrary  constant $\lambda > 0$, have 
seed length $O(\log^2 n)$, and their output
length is $\approx n^{\lambda/3}$. 
\smallskip

\section{Introduction}
Extractors are procedures that remedy an imperfect source of random strings. They have been the object of
intense research in the last years and several relevant techniques have been developed. This paper puts forward  a new framework for constructing extractors based 
on a new connection between extractors and pseudo-random generators. 
Surely, in some regards, there are obvious similarities between the two concepts.
 A pseudo-random generator  takes as input a short random string 
called the seed and outputs a long string
that cannot be distinguished from a truly random string by any test that is computable by circuits of bounded size.
An extractor has two inputs: (a) The first one comes from an imperfect (\ie, with biased bits and correlations among bits) 
distribution on binary strings of some length and it is called the weakly-random string; (b)
the second one is a short random seed. The output is a long string that cannot be distinguished from a truly random string by
\emph{any} test. 
One difference between pseudo-random generators and extractors is the number of inputs (one versus two). 
From a technical point of
view this difference is minor because the known constructions of pseudo-random generators 
implicitly do use an extra input which is a function that
in some sense is computationally hard. 
The fundamental difference is in the randomness requirement for the output. Thus, while
the output of a pseudo-random generator looks random in a complexity-theoretic way, the output of an extractor is random 
(or very close to random) in an absolute information-theoretic way. Consequently pseudo-random generators and 
extractors appear to 
belong to two very different worlds, and, for many years, the developments in the construction of pseudo-random 
generators and extractors went
along distinct research lines.

Trevisan~\cite{tre:j:extract} has made a breakthrough contribution in this area by observing that 
the (apparently superficial) similarity between extractors and pseudo-random generators 
extends to some of the methods to build the two kind of objects. For the reasons mentioned above, Trevisan's result has been
extremely surprising. It has also been an isolated
example of a transfer from the complexity theory standard arsenal of techniques to the information
theoretical area. In this paper we extend Trevisan's observation and establish that, as far as construction methods 
are concerned, 
there is a truly close relationship 
between pseudo-random generators
and extractors. Specifically, we show that the other major route (than the one followed by Trevisan) 
that leads to
pseudo-random generators (of a 
somewhat different kind) can also be used to construct extractors. Some explanations are in order at this point.

There are two known approaches for constructing pseudo-random generators. One approach  uses as a
building block a hard function $f$ and, in one typical setting of parameters, for any given $k \in \nat$, builds
a pseudo-random generator $g$ with outputs of length $n$ that is secure against adversary tests computable in time $n^k$. 
The running
time to compute $g(x)$ is $n^{k'}$, for some $k' > k$. This kind of pseudo-random generators can be used for 
derandomizing $\bpp$ 
computations. They cannot be used in cryptography, because in this setting, it is unwise to assume that the adversary
is endowed with less computational power ($n^k$) than the legitimate users ($n^{k'}$). Henceforth we will call this
type of pseudo-random generator  a ``derandomization pseudo-random generator'' 
(also known as a Nisan-Wigderson pseudo-random generator).

The second approach uses as a building block a hard object of a more sophisticated type, namely a one-way function
(the hardness of such a function $f$ consists in the difficulty to invert it, but $f$ must satisfy an additional
property, namely, it should be easy to calculate $f(x)$ given $x$). It is known that given a one-way function,
one can construct a pseudo-random generator~\cite{hasimplevlub:j:psgenoneway}. An easier construction produces a pseudo-random generator from any
one-way length-preserving permutation. This second approach has the disadvantage that is using as a building block
a more demanding type of object. The advantage of the method is that a pseudo-random generator $g$ constructed in this way can be used
in cryptography because $g(x)$ can be calculated in time significantly shorter than the time an adversary must spend
to distinguish $g(x)$ from a truly random string. Henceforth we will call this
type of pseudo-random generator a ``crypto pseudo-random generator'' 
(also known as a Blum-Micali-Yao  pseudo-random generator).

Trevisan has shown that the known methods for constructing  derandomization pseudo-random generators also produce extractors. 
More precisely, he has shown that the constructions of pseudo-random generators from hard functions given by
Nisan and Wigderson~\cite{nis-wig:j:hard} and Impagliazzo and Wigderson~\cite{imp-wig:c:pbpp} can be used almost
directly to produce extractors. His method has been extended in a number of papers to build extractors with
increasingly better parameters (see the survey paper by Shaltiel~\cite{sha:j:extract}).
In the 
paper~\cite{tre:c:extractor}, the conference version of~\cite{tre:j:extract}, Trevisan has suggested that 
the methods
to construct crypto pseudo-random generator cannot be used to build extractors. 
We show that in fact they can, at least for a combination of parameters that, even though not optimal, is not trivial. 
Moreover,
we show that the extractors constructed in this way are very simple and efficient.

An extractor can be viewed as a bipartite graph and is therefore a static finite object that can
be constructed trivially by exhaustive search. We are looking however for efficient constructions. 
Typically ``efficient"  means ``polynomial time," but one can envision different levels of efficiency 
and one remarkable such level would certainly be ``linear time." 
The first extractor built in this paper  follows almost directly the classical construction of a pseudo-random generator
 from a one-way permutation, and comes very close to this level of efficiency: 
Viewed as a procedure, it runs in $O(n \log n)$ time (in the standard RAM model). 
In addition it is very simple. The following is a complete description of it.  
The input consists of the weakly-random string $X$, of length $n = \tn 2^{\tn}$ for some 
integer $\tn$, and of the seed $((x_1, \ldots, x_\ell), r)$, with $|x_i| = \tn$, $\ell = O(\tn)$, and $|r| = \ell \tn$. 
We view $X$ as a function $X: \zo^{\tn} \mapping \zo^{\tn}$, and, using the standard procedure, we transform $X$ 
into a circular permutation $R: \zo^{\tn} \mapping \zo^{\tn}$. For $i = 0$ to $m-1 = n^{\Omega(1)}$, we calculate $b_i$ as 
the inner product modulo $2$ of $r$ and $(R^i(x_1) \ldots R^i(x_\ell))$. The output is $b_0 \ldots b_{m-1}$.

Another remarkable level of efficiency which has received a lot of attention recently is that of sublinear time. It
may be the case that in some applications we only need the $i$-th bit from the sequence of random bits that are
extracted from the weakly-random string. We would like to obtain this bit in time polynomial in the length of the
index $i$, which typically means polylog time in the input length (under the assumption that each input bit can be
accessed in one time unit). By analogy with the case of list-decodable codes, we call an extractor with this property,
a \emph{bitwise locally computable extractor.}\footnote{The simpler name \emph{locally computable extractor} is already
taken by a different kind of efficient extractors, namely by extractors computable in space linear in the output length,
see~\cite{vad:j:localextractor}, \cite{lu:j:localextractor}.} The second extractor that we build is of this type. The
algorithm deviates from the direct construction of a pseudo-random generator from a one-way function. However it
relies on a basic idea used in the construction of the first extractor, combined with the idea of taking
consecutive inputs of the hard function as in the extractor of Ta-Shma, Zuckerman and 
Safra~\cite{ta-shma-zuc-saf:c:extractor}.
This second extractor is even simpler and its complete description is as follows. The input consists of the weakly-random
string $X$ of length $n = \tn \cdot 2^{\tn}$, for some natural number $\tn$, and of the seed $((x_1, \ldots, x_\ell), r)$,
with $|x_i| = \tn$, for all $i$, $\ell = O(\tn)$, and $|r| = \ell \tn$. We view $X$ as the truth-table of a function
$X: \zo^{\tn} \mapping \zo^{\tn}$. For $i = 0$ to $m-1 = n^{\Omega(1)}$, we calculate $b_i$ as the inner product modulo 2 of $r$ and
$(X(x_1 + i), \ldots, X(x_\ell + i))$, where the addition is done modulo $2^{\tn}$. The output is $b_0 \ldots b_{m-1}$.

The parameters of the extractors constructed in this paper are not optimal. Both extractors that have been described above 
work for weak sources having min-entropy $\lambda n$, for arbitrary constant $\lambda > 0$, use a random seed of length 
$O(\log^2 n)$, and the output length is approximately $n^{\lambda/3}$. A variant of the second extractor has seed length
$O(\log n)$ (here, for simplicity, we assume that the extractor's error parameter $\epsilon$ is a constant), but the output length reduces  to $2^{O(\sqrt{\log n})}$.

Lu's extractor~\cite{lu:j:localextractor} coupled with the constructions of designs from the paper of Hartman 
and Raz~\cite{har-raz:j:design} can be seen to be also a 
bitwise locally computable extractor with parameters similar to those of our second extractor (note that the designs in~\cite{har-raz:j:design} appear to imply extractors with seed length $\Omega(\log^2 n)$). Lu's extractor is using 
expander graphs and the designs from~\cite{har-raz:j:design} need somewhat unwieldy algebraic objects. 
It seems to us that the extractors presented in this paper are simpler than all the extractors from the
literature.\footnote{We note that Dziembowski and Maurer~\cite{dm:j:simpleext} give a similarly simple construction of
an object that is related to extractors.} 
At the highest
level of abstraction, our extractors follow the ``reconstruction paradigm'' (see~\cite{sha:j:extract}) typical to
Trevisan's extractor and to its improvements~\cite{rareva:c:extractor, ta-shma-zuc-saf:c:extractor, sha-um:c:extractor}. 
The major differences are that our extractors
avoid (1) the use of designs (in this respect they are similar to the extractors in~\cite{ta-shma-zuc-saf:c:extractor} 
and~\cite{sha-um:c:extractor}),
and, perhaps more strikingly, (2) the encoding of the weakly-random string with an error-correcting code
having a good list-decoding property. Our extractors can be implemented very easily and are thus suitable for
practical applications. For example, they can be utilized to generate one-time pad keys in cryptosystems based on the
bounded-storage model (see the papers of Lu~\cite{lu:j:localextractor} and Vadhan~\cite{vad:j:localextractor}), or for constructions of error-correcting codes using the scheme in~\cite{ta-shma-zuc:c:extractorcodes} (the extractors built in this paper are actually strong extractors---for definition see, for example~\cite{sha:j:extract}---as required by this scheme). They may also
have theoretical applications in situations where the kind of efficiency achieved by our extractors is essential.

\section{Definitions}
Notations: $x \odot y$ denotes the concatenation of the strings $x$ and $y$, $|x|$ denotes the length of the string $x$, and
$\lll A \rrr$ denotes the cardinality of the set $A$. We remind the standard definition of an extractor. Let $n \in \nat$. Let $X_n$, $Y_n$ be two distributions on~$\sigman$. The statistical distance between $X_n$ and $Y_n$ is denoted
$\Delta_{{\rm stat}}(X_n, Y_n)$ and is defined by $\Delta_{{\rm stat}}(X_n, Y_n) = \max_{A \subseteq \zon} | \prob (X_n \in A) - \prob(Y_n \in A) |$.

If we view the sets $A \subseteq \zon$ as \emph{statistical tests}, then, by the above expression, 
$\Delta_{{\rm stat}}(X_n, Y_n) \leq \epsilon$ signifies that no test can distinguish between the
distributions $X_n$ and $Y_n$ except with a small bias $\epsilon$. If we restrict to tests that can be calculated
by bounded circuits, we obtain the notion of computational distance between distributions. Namely, the 
computational distance between $X_n$ and $Y_n$ relative to size $S$ is denoted
$\Delta_{{\rm comp}, S}(X_n, Y_n)$ and is defined by
$\Delta_{{\rm comp}, S}(X_n, Y_n) = \max \left | \prob (C(X_n) = 1) - \prob (C(Y_n) = 1) \right |$, 
where the maximum is taken over all circuits $C$ of size $\leq S$. Abusing notation, we identify a circuit $C$ with the set of strings $x$ for which $C(x) = 1$. Thus, $x \in C$ is equivalent to $C(x)=1$.

The min-entropy of a distribution is a good indicator of the degree of randomness of the distribution. 
The min-entropy of a random variable taking values in
$\zon$ is given by
$\min \Big\{ \log \frac{1}{\prob(X=a)} ~ \Big |~ a \in \zon, \prob(X=a) \not= 0 \Big\}$.

Thus if $X$ has min-entropy $\geq k$, then for all $a$ in the range of $X$, ${\prob(X=a) \leq 1/2^k}$. 
For each $n \in \nat$, let $U_n$ denote the uniform distribution over $\zon$. 
We are now ready to define an extractor formally.
\begin{definition}
\emph{(Extractor)}
The values $n, k, d, m$ are integer parameters, and $\epsilon > 0$ is a real number parameter. A function
$E\colon \zon \times \zo^d \mapping \zo^m$ is a $(k, \epsilon)$-extractor if for every distribution
$X$ on $\zon$ with min-entropy at least $k$, the distribution $E(X, U_d)$ is $\epsilon$-close to the
uniform distribution $U_m$ in the statistical sense, \ie,
$\Delta_{{\rm stat}}(E(X, U_d), U_m) \leq \epsilon.$
\end{definition}
Thus, an extractor has as input (a) a string $x$ produced by an imperfect source with distribution $X$, where the defect of the
distribution is measured by $k = \mbox{min-entropy}(X)$, and (b) a random seed $y$ of length $d$. The output is
$E(x,y)$, a string of length $m$. The key property is that, for every subset $W \subseteq \Sigma^m$,
\begin{equation}
\label{e:extract1}
\lvert \prob_{x \in_X \zon, y \in \zo^d} (E(x,y) \in W) - \prob_{z \in \Sigma^m} (z \in W) \rvert \leq \epsilon.
\end{equation}

If we consider $n$ and $k$ as given
(these are the parameters of the source), it is desirable that $d$ is small, $m$ is large, and $\epsilon$ is small. It can be shown nonconstructively that for every $k \leq n$ and $\epsilon > 0$, there
exist extractors with $d = \log (n-k) + 2 \log(1/\epsilon) + O(1)$ and 
$m = k + d - 2 \log(1/\epsilon) - O(1)$. It has been shown~\cite{rad-tas:j:extractors} that these parameters are optimal.
Furthermore, we want the family of extractors to be efficiently computable. For simplicity, we have defined individual 
extractors. However, implicitely we think of a family of extractors indexed
by $n$ and with the other parameters being uniform functions of $n$. In this way we can talk about efficient constructions
of extractors by looking at the time and space required to calculate $E(x,y)$ as functions of $n$.

An extractor $E\colon \zo^n \times \zo^d \mapping \zo^m$ can also be viewed as a regular bipartite graph where
the set of ``left'' nodes is $V_{\text{left}} = \zo^n$ and the set of ``right'' nodes  is $V_{\text{right}} = \zo^m$.
The degree of each node in $V_{\text{left}}$ is $2^d$, and two nodes $x \in V_{\text{left}}$ and $z \in V_{\text{right}}$
are connected if there is $y \in \zo^d$ such that $E(x,y) = z$. We can imagine that each $x \in V_{\text{left}} = \zo^n$
is throwing $2^d$ arrows at $V_{\text{right}} = \zo^m$.

To understand better Equation~(\ref{e:extract1}), let us look deeper into the structure of an extractor. 
We fix parameters $n, d, m$ and $\epsilon$ and a function ${E\colon \zo^n \times \zo^d \mapping \zo^m}$.  Let us consider
an arbitrary set $W \subseteq \zo^m$ and a string $x \in \zon$. We say that $x$ hits $W$ $\epsilon$-correctly via $E$ if
the fraction of outgoing edges from $x$ that land in $W$ is $\epsilon$-close to the fraction 
$\lVert W \rVert / \lVert \zo^m \rVert$, \ie,
\[
\biggl\lvert \frac{ \lVert \{ E(x, y) \mid y \in \zo^d \} \cap W \rVert} { \lVert \zo^d \rVert} -
\frac{\lVert W \rVert}{\lVert \zo^m \rVert} \biggr \rvert \leq \epsilon.
\]
If we look at a fixed $x$, it cannot hold that for every $W \subseteq \zo^m$, $x$ hits $W$ $\epsilon$-correctly
(for example, take $W = \{E(x, y) \mid y \in \zo^d \}$). Fortunately, for $E$ to be an extractor, all we need is that
any $W \subseteq \zo^m$ is hit $\epsilon$-correctly by most $x \in \zon$. The folowing lemma has appeared more or less
explicitly in the literature (see, for example,~\cite{sha:j:extract}).
\begin{lemma}
\label{l:extractor1}
Let $E\colon \zon \times \zo^d \mapping \zo^m$ and $\epsilon >0$. Suppose that for every $W \subseteq \zo^m$, 
the number of $x \in \zon$ that
do not hit $W$ $\epsilon$-correctly via $E$ is at most $2^t$, for some $t$. 
Then $E$ is a $(t + \log (1/\epsilon), 2 \epsilon)$-extractor.  
\end{lemma}
$\proof$
Let $X$ be a distribution on $\zon$ with min-entropy at least $t + \log (1/\epsilon)$ and let $W$ be
a subset of $\zo^m$. There are at most $2^t$ $x$'s that do not hit $W$ $\epsilon$-correctly and the 
distribution $X$ allocates to these $x$'s a mass probability of at most $2^t \cdot 2^{- (t + \log (1/\epsilon))} = \epsilon$.
We have,
\begin{equation*}
\begin{array}{rll} 
\prob_{x \in_X \zon, y \in \zo^d} ( E(x,y) \in W) & &\\  
& \mspace{-150mu} = & \mspace{-150mu} \quad\prob_{x \in_X \zon, y \in \zo^d} ( \mbox{$E(x,y) \in W$ and $x$ hits $W$ $\epsilon$-correctly}) \\ 
& \mspace{-150mu} + &  \mspace{-150mu} \quad~~~\prob_{x \in_X \zon, y \in \zo^d} ( \mbox{$E(x,y) \in W$ and $x$ does not hit $W$ $\epsilon$-correctly}).
\end{array}
\end{equation*}
The first term in the right hand side is between $\frac{ \lll W \rrr}{\lll \zo^m \rrr}- \epsilon$ and 
 $\frac{ \lll W \rrr}{\lll \zo^m \rrr} + \epsilon$, because for each $x$ that hits $W$ $\epsilon$-correctly,
\[
\prob_{y \in \zo^d}(E(x,y) \in W) \in \left[ \frac{ \lll W \rrr}{\lll \zo^m \rrr}- \epsilon, 
\frac{ \lll W \rrr}{\lll \zo^m \rrr} + \epsilon \right].
\]
The second term is bounded by 
$$\prob_{x \in_X \zon, y \in \zo^d} ( \mbox{$x$ does not hit $W$ $\epsilon$-correctly}),$$ 
which is, as we have seen, between $0$ and $\epsilon$. Plugging these estimates in the above equation, 
we obtain that
\[
\left \lvert 
\prob_{x \in_X \zon, y \in zo^d}( E(x,y) \in W) - \frac{ \lll W \rrr}{\lll \zo^m \rrr} \right \rvert \leq 2\epsilon.
\]
Thus, $E$ is a $(t + \log \frac{1}{\epsilon}, 2 \epsilon)$-extractor.~\qed

We recall the definition of a pseudo-random generator.
\begin{definition}
\label{d:prgen}
\emph{(Pseudo-random generator)}\index{pseudo-random generator}
Let $\ell, L, S \in \nat$ and $\epsilon > 0$ be parameters. A function $g\colon \sigmal \mapping \Sigma^L$ is
a pseudo-random generator with security $(\epsilon, S)$ if 
$\Delta_{{\rm comp}, S}(g(U_\ell), U_L) \leq \epsilon$.
\end{definition}

\section{Overview and comparison with Trevisan's approach}

Trevisan's method is based on the constructions of pseudo-random generators from hard functions given 
in~\cite{nis-wig:j:hard} and in~\cite{imp-wig:c:pbpp}. These constructions use a function $f$ as a block-box and
construct from it a function $g_f$ that stretches the input (\ie, $|g_f(x)| >> |x|$) and which has the following property.
If there exists a circuit $D$ that distinguishes $g_f(x)$, when $x$ is randomly chosen in the domain of $g_f$, from the
uniform distribution, then there is a small circuit $A$, which uses $D$ as a subroutine, such that $A$ calculates $f$ 
(or an approximation of $f$, depending on whether we are using the method in~\cite{imp-wig:c:pbpp} or the one
in~\cite{nis-wig:j:hard}). Therefore if $f$ is a hard function, there can be no circuit $D$ as above of small size and
thus $g_f$ is a pseudo-random generator. Trevisan has observed that (1) the truth-table of $f$ can be viewed as a string
produced by a weak source that can serve as an extra input of the pseudo-random generator, 
and (2) the circuit $A$ invoking $D$ can be
considered as a special type of a circuit that is endowed with $D$-gates. By a standard counting argument, it can be 
shown that, for any circuit $D$, regardless of its size, the set of functions
that can be calculated by small circuits with $D$-gates is small. A circuit $D$ can be viewed statically as
a statistical test (more exactly, the statistical test associated to the circuit $D$ is the set of strings accepted 
by $D$). In the new terminology, the fact that $D$ distinguishes the distribution of $g_f(x)$ from the uniform 
distribution with $\epsilon$ bias can be restated as ``$f$ does not hit $D$ $\epsilon$-correctly via $g$.'' 
The main property mentioned above can be restated as saying that the set of functions $f$ that do not hit
$D$ $\epsilon$-correctly is included in the set of functions computable by small circuits with $D$-gates. Since the
latter set is small, the former set is small as well, and thus,
by Lemma~\ref{l:extractor1}, the construction yields an extractor. In a nutshell, Trevisan's method 
replaces hard functions
(a complexity-theoretic concept) with random functions (an information-theoretic concept) and takes advantage of the 
fact that a random function is hard and thus the construction carries over in the new setting.

We would like to follow a similar approach for the construction of crypto pseudo-random generators from one-way permutations.
Those constructions do use a one-way permutation $R$ as a black box to construct a pseudo-random generator $g_R$, 
and thus a truth-table of $R$ can be considered as an extra input of the pseudo-random generator. 
Also, the proof is a reduction that shows that
if a circuit $D$ distinguishes $g_R(x)$ from the uniform distribution, then there is a small circuit $A$, 
invoking the circuit $D$, that inverts $R$ on a large fraction of inputs. 
To close the proof in a similar way to Trevisan's approach, we would need to argue 
that the vast majority of permutations are one-way. It seems that we hit a major obstacle because, unlike the 
case of hard functions, it is not currently known if even a single one-way function exists (and we are seeking 
an unconditional proof for the extractors that we build). We go around this obstacle by allowing algorithms to have
oracle access to the function they compute. Thus, in the above analysis, the circuit $A$, in addition to invoking
the circuit $D$, will also have oracle access to the permutation $R$. In this setting all permutations are easy to
compute because,  obviously, 
 there is a trivial constant-time  algorithm that, for any permutation $R:\zon \mapping \zon$,
given the possibility to query $R$, calculates $R(x)$. We need to argue that only few permutations $R$ are invertible by 
algorithms that can query $R$ in a bounded fashion.  More precisely we need to estimate the
size of the set of permutations $R: \zon \mapping \zon$ that can be inverted on a set of $T$ elements in $\zon$ by
circuits that can pose $Q$ queries to $R$. This problem has been considered by 
Impagliazzo~\cite{imp:m:randomoneway} and by Gennaro and Trevisan~\cite{gen-tre:c:cryptlowerbounds}. Their techniques
seem to work for the case $T \cdot Q < 2^n$ and lead to extractors that work only for sources with high 
min-entropy.\footnote{On the other hand, these extractors have the interesting property that their output looks
random even to statistical tests that have some type of access to the weakly-random string. These results will be reported 
in a separate paper.}

We obtain better parameters by restricting the type of one-way permutations and the type of circuits that attempt to
invert them. A second look at the standard construction of Blum-Micali-Yao pseudo-random generators reveals that the
circuit $A$ with $D$-gates manages to determine $x$  using only the values $R(x), R^2(x), \ldots,
R^m(x)$ (where $m$ is the generator's output length). It is thus enough to consider only circuits that use
this pattern of queries to the permutation $R$. Intuitively, for a random permutation $R$, the value of $x$ should be
almost independent of the values of $R(x), R^2(x), \ldots,
R^m(x)$, and thus, a circuit $A$ restricted as above cannot invert but a very small fraction of permutations. If we
take $R$ to be a random circular permutation, the above intuition  can be easily turned into a proof 
based on a Kolmogorov-complexity 
counting argument. A circular permutation $R: \zon \mapping \zon$ is fully specified by the sequence
$(R(1), R^2(1), \ldots, R^{N-1}(1))$, where $N = 2^n$. If a circuit $A$ restricted as above inverts $R(x)$ for all $x$,
then the permutation $R$ is determined by the last $m$ values in the above sequence, namely $R^{N-m}(1), 
R^{N-(m-1)}(1), \ldots, R^{N-1}(1)$. Indeed, given the above values, the circuit $A$ can determine
$R^{N-m-1}(1)$, which is $R^{-1}(R^{N-m}(1))$, and then $R^{N-m-2}(1)$, and so on till $R(1)$ is determined. Therefore
such a permutation $R$, given the circuit $A$, can be described concisely using only  $m\cdot n$ bits (for 
specifying, as discussed, the last $m$ elements in the above sequence). In fact, in our case, 
the circuit $A$ does not invert $R(x)$ 
for all $x \in \zon$, and, therefore, the values of $R$ at the points where the inversion fails have to be
included in the description. A further complication is that even for the successful cases, the circuit $A$ only
list-inverts $R(x)$, which means that $A$ on input $R(x)$ produces a relatively short list of elements, one of which is
$x$. Thus, one also has to include in the description of $R$ the rank of $x$ in the list produced by $A$. The
quantitative analysis of the standard construction of a crypto pseudo-random generator shows that if the
permutation $R$ does not hit $D$ $\epsilon$-correctly, then the circuit $A$ with $D$-gates is only able to
produce for an $\epsilon/m$ fraction of $R(x), x \in \zon$, a list with $m^2/\epsilon^2$ elements one of which is $x$. For
interesting values of $m$ (the pseudo generator's output length), the $\epsilon/m$ fraction is too small and
needs to be amplified to a value of the form $(1-\delta)$, for a small constant $\delta$. This can be done by
employing another technique that is well-known in the context of one-way functions. Namely, we use Yao's method
of converting a weak one-way function into a strong one-way function by taking the direct product. In other words,
we start with a circular permutation $R$, define (the direct product) $\barR(x_1, \ldots, x_\ell) = R(x_1) \odot 
\ldots \odot R(x_\ell)$ (where $\odot$ denotes concatenation), for some appropriate value of $\ell$, and use
$\barR$ in the definition of the extractor (instead of $R$ in our tentative plan sketched above).
It can be shown that, for $\ell = O((1/\delta) \log (1/\gamma))$, if a circuit $A$ list-inverts $(y_1, \ldots, y_\ell)$, with list size $T = m^2/\epsilon^2$, for
a $\gamma = \epsilon/m$ fraction of $\ell$-tuples $(y_1, \ldots, y_\ell) \in (\zon)^\ell$, then there is a
probabilistic algorithm $A'$ that list-inverts $R(x)$ with list size $O(n \cdot T \cdot (1/\delta) \cdot (1/\gamma) 
\cdot \log(1/\gamma))$ for a $(1-\delta)$ fraction of $x \in \zon$. By fixing the random bits and the
queries that depend on these random bits, we can obtain a brief description of $R$ as in our first tentative plan. It
follows that only few permutations $R$ can hit $D$ $\epsilon$-incorrectly and, therefore, by Lemma~\ref{l:extractor1},
we have almost obtained an extractor (we also need to convert an arbitrary function $X:\zon \mapping \zon$ into a
circular permutation $R:\zon \mapping \zon$, which is an easy task).

Briefly, the proof relies on the fact that if a permutation $R$ does not hit $D$ $\epsilon$-correctly, then there must be
a very strong dependency between the ``consecutive'' values $x, R(x), R^2(x), \ldots, R^m(x)$, for many $x \in \zon$, and
only few permutations $R$ exhibit such dependencies.

The second extractor starts from this idea and the observation that, for the sake of building an extractor, we can
work with a function $\barX$ (\ie, not necessarily a permutation) and consider consecutive values
$\barX(\barx), \barX(\barx+1), \ldots, \barX(\barx+m)$, as in the extractor of Ta-Shma, Zuckerman, and 
Safra~\cite{ta-shma-zuc-saf:c:extractor}.
That extractor (as well as all the extractors using the ``reconstruction paradigm'') takes $\barX$ to be the
encoding of an arbitrary function $X$ with a good list-decoding property and some other special algebraic properties.
This is necessary, among other things, for the same type of amplification as in our discussion above. We use instead 
a direct-product construction that is much simpler to implement (however, the cost is a longer seed length).
\section{Restricted permutations, restricted circuits}
\label{s:randperm}
The space from where we randomly choose permutations consists of permutations of a special form. 
First we consider the set $\crc$ of all circular permutations $R: \zon \mapping \zon$. Next, for some parameter
$\ell \in \nat$, we take the $\ell$-direct product of $\crc$. This means that for any $R \in \crc$, we define
$\barR_\ell: \zo^{\ell n} \mapping \zo^{\ell n}$ by $\barR_\ell (x_1 \odot x_2 \odot \ldots \odot x_\ell) = R(x_1) \odot R(x_2) \odot
\ldots \odot R(x_\ell)$. We let $\perm_\ell$ be the set $\{\barR_\ell \mid R \in \crc \}$. We will drop the subscript
$\ell$ when its value is clear from the context or when it is not relevant in the discussion.

We want to argue that no circuit that queries $\barR$ in a restricted way can invert a ``large'' fraction of
$\barR(\barx)$ except for a ``small'' fraction of permutations $\barR$ in $\perm$. In order to obtain adequate values
for ``large'' and ``small'' we will impose the following restriction on the pattern of queries that the circuit can make.
\begin{definition}
An oracle circuit $C$ on inputs of length at least $\ell \cdot n$ is $L$-restricted if on any input $x$ and for all oracles
$\barR \in \perm_\ell$, $C$ only queries $x_{\rm first}, \barR(x_{\rm first}), \barR^2(x_{\rm first}), \ldots,
\barR^{L-1}(x_{\rm first})$, where $x_{\rm first}$ is the string consisting of the first $\ell \cdot n$ bits of $x$.
\end{definition}
We will allow the circuits to attempt to invert $\barR$ in a weaker form: On input $\barR(\barx)$, $C^{\barR}$ outputs a 
small list of strings one of which (in case $C$ succeeds) is $\barx$. When this event happens, we say that $C^{\barR}$
\emph{list-inverts} $\barx$. 
We are interested in estimating the number
of permutations $\barR \in \perm$ so that $C^{\barR}$ list-inverts $\barR(\barx)$ for a large fraction of $\barx$.
\begin{definition}
\label{d:goodperm}
Let $C$ be an oracle circuit. A permutation $\barR$ is $(\gamma, T)$-good for $C$ if for at least a
$\gamma$ fraction of $\barx \in \zo^{\ell n}$, $C^{\barR}$ on input $\barR(\barx)$ outputs a list of $T$ elements 
that contains $\barx$. 
\end{definition}
We will show that a permutation that is $(\gamma, T)$-good for a restricted circuit $C$ admits
a short description conditioned by $C$ being given. This leads immediately to an estimation of the number of
permutations $\barR$ that are $(\gamma, T)$-good for a given restricted circuit $C$.
\begin{lemma}
\label{l:description}
Let $\gamma > 0$, $n \in \nat, L \in \nat$, and $T \in \nat$. Let $N = 2^n$. 
Let $\delta > 0$ and let $\ell = \big \lceil \frac{3}{\delta} \cdot \log\big( \frac{2}{\gamma}\big)\big \rceil$.
Assume $\delta \geq 2e^{-n}$ and $\ell < L+1$.
Let $C$ be an $L$-restricted circuit, having inputs of length $\ell n$, 
and let $\barR \in \perm_{\ell}$ be a permutation that is $(\gamma, T)$-good for $C$.
Then, given $C$ and $\ell$,  $\barR$ can be described using a number of bits that is bounded by
$2\delta N n + Ln + N \log n + (\log 6)N + N \log (1/\delta) + N \log \log (2/\gamma) + N \log (1/\gamma) 
+ N \log T  + 18n^2 \cdot L \cdot \frac{1}{\gamma} \cdot \big( \frac{1}{\delta} \big)^2 
\big( \log \frac{2}{\gamma} \big)^2$.
\end{lemma}
$\proof$
Since $\barR$ is the $\ell$-direct product of $R$, it is enough to present a short description of $R$. We will
first show that the assumption that $C^{\barR}$ list-inverts a $\gamma$-fraction of $\barR(\barx)$ with $\barx \in \zo^{\ell n}$ implies
that there exists an oracle circuit $B$ so that $B^R$ list-inverts a $(1-\delta)$ fraction of $R(x)$ with $x \in \zo^n$. The
circuit $B$ is not $L$-restricted but it has a similar property. Namely the circuit $B$ makes two categories of
queries to the oracle $R$. The first category consists of a set of queries that do not depend on the input. The second
category depends on the input $y$ and it  consists of the queries $y, R(y), \ldots, R^{L-1}(y)$. The circuit $B$ is
helpful in producing the concise description of $R$ that we are seeking. Note that the permutation
$R \in \crc$ is determined by the vector $(R(1), R^2(1), \ldots, R^{(N-1)}(1) )$. This vector will be described in the
following way. The last $L$ entries are described by themselves. Then we describe each of the other entries $y$ one at
a time going backwards in the vector. Suppose that $R(y), R^2(y), \ldots, R^{L}(y)$ are already described. We describe now
the preceding term in the sequence, which is $y$. There are two cases. 

\emph{Case 1}: $B^R$ list-inverts $R(y)$. In this
case $y$ is determined by its rank in the $T$-list produced by $B^R$ on input $R(y)$. The computation of $B^R$ on input
$R(y)$ depends on the strings $R(y), \ldots, R^L(y)$ (which are already described) and on 
the value of $R$ on the fixed queries (these values 
have to be given in the description, but they are common to all the entries in the vector). 

\emph{Case 2}: If $B^R$ fails
to list-invert $y$ (this will happen only for a small fraction $\delta$ of $y$'s), then $y$ is described by itself. 

We will
show that this description policy needs the asserted number of bits.

We proceed with the technical details. The amplification of the fraction of inverted inputs from $\gamma$ to
$(1-\delta)$ is done using the well-known technique of producing strong one-way functions from weak one-way functions
(Yao~\cite{yao:c:oneway}).
Let $w = 6 \cdot \frac{1}{\delta} \cdot \log \big(2/\gamma\big) \cdot \frac{1}{\gamma}$. 
Recall that $\ell = \big \lceil \frac{3}{\delta} \cdot \log\big( \frac{2}{\gamma}\big) \big \rceil$. 
It holds that, for $n \geq \ln(2/\delta)$ and $\delta < 1/3$, 
$\frac{\ell}{w} < \gamma - (1-\delta + e^{-n})^\ell$.
Let ${\rm{INV}}$ be the set of strings $\barR(\barx)$ on which $C^{\barR}$ outputs a $T$-list that contains
$\barx$. From the hypothesis, we know that $\lll {\rm{INV}} \rrr \geq \gamma \cdot 2^{\ell n}$. 
We define the following probabilistic algorithm  $D$.

\myrule

Input: $y = R(x)$, for some $x \in \sigman$. Goal: Find a short list that contains $x$.

${\rm LIST} = \emptyset.$

Repeat the following $n \cdot w$ times.

~~~~~~~~Pick random $i \in \{1, \ldots, \ell\}$.

~~~~~~~~Pick $\ell-1$ random strings in $\zon$ denoted $y_{1}, \ldots, y_{i-1}, y_{i+1}, \ldots, y_\ell$.

~~~~~~~~Calculate $Y = y_1 \odot \ldots \odot y_{i-1} \odot R(x) \odot y_{i+1} \odot \ldots \odot y_\ell$.

~~~~~~~~Call the circuit $C^{\barR}$ to invert $Y$. $C^{\barR}$ returns a $T$-list of $\ell$-tuples in $(\zon)^\ell$.

~~~~~~~~(Note: In case of success one of these $\ell$-tuples is 

~~~~~~~~$(R^{-1}(y_1), \ldots, R^{-1}(y_{i-1}), x, R^{-1}(y_{i-1}), \ldots, R^{-1}(y_{\ell}))$.)

~~~~~~~~Add to ${\rm LIST}$ the $i$-th component of every $\ell$-tuple in the list produced by $C^{\barR}$.

End Repeat

\myrule

We say that the above algorithm is successful on input $y=R(x)$ if, at the conclusion of the algorithm, 
${\rm LIST}$ contains $x$. 
We estimate the success probability of the above circuit on input $y=R(x)$. 

Let $N(y)$ be the multiset of $\ell$-tuples having $y$ as one component where the multiplicity of a tuple is the number of
occurrences of $y$ in the tuple. For a set $A \subseteq \zon$, we define $N(A) = \bigcup_{y \in A} N(y)$. 
It can be seen that, for all $y \in \zon$,
$\lll N(y) \rrr = \ell \cdot 2^{n(\ell-1)}$.
We define
\[
V_w = \Big \{y \in \zon \mid \frac{\lll N(y) \cap {\rm INV} \rrr}{\lll N(y) \rrr} \geq \frac{1}{w} \Big \}.
\]
Let $\overline{V_w}$ be the complement of $V_w$. We have
\begin{eqnarray*}
\lll N(\overline{V_w}) \cap {\rm{INV}} \rrr & \leq & \sum_{y \in \overline{V_w}} \lll N(y) \cap {\rm INV} \rrr \\
& < & 2^n \cdot \frac{1}{w} \cdot( \ell \cdot 2^{n(\ell-1)}) \\
& = & \frac{\ell}{w} \cdot \lll (\sigman)^\ell \rrr.
\end{eqnarray*} 
We show that this is possible only if $\lll \overline{V_w} \rrr < (\delta -e^{-n})\cdot \lll \sigman \rrr$.
Let $A \subseteq \sigman$ be a set with $\lll A \rrr  \geq (\delta - e^{_n}) \cdot \lll \sigman \rrr$.
We observe that
$N(A)$ covers an overwhelming fraction of $(\sigman)^\ell$. Indeed, note that the probability that a tuple
$(y_1, \ldots, y_\ell)$ is not in $N(A)$ is equal to the probability of the event 
``${y_1 \not\in A~\AND \ldots \AND~y_\ell \not\in A}$'' which is bounded by $(1-\delta + e^{-n})^\ell$.
Therefore, the complementary set of $N(A)$, denoted $\overline{N(A)}$, satisfies
\[
\lll \overline{N(A)} \rrr < (1-\delta + e^{-n})^\ell \cdot \lll (\sigman)^\ell \rrr.
\]
Then,
\begin{eqnarray*}
\lll N(A) \cap {\rm{INV}} \rrr & = & \lll {\rm{INV}} \rrr - \lll {\rm{INV}} \cap  \overline{N(A)} \rrr \\
& \geq & \lll {\rm{INV}} \rrr - \lll \overline{N(A)} \rrr \\
& > & \big[ \gamma - (1-\delta + e^{-n})^\ell \big]\lll (\sigman)^\ell \rrr.
\end{eqnarray*}
Recall that
$\ell/w  < \big [ \gamma - (1-\delta + e^{-n})^\ell \big ]$.
Thus necessarily $\lll \overline{V_w} \rrr < (\delta - e^{-n}) \cdot 2^n$.

On input $y = R(x)$, at each iteration, the algorithm chooses
uniformly at random $\overline{y}$ in $N(y)$. The circuit $C$ is invoked next to invert $\overline{R}(\overline{y})$.
The algorithm succeeds if and only if $\overline{y} \in {\rm INV}$. For all $y \in V_w$, 
$\frac{\lll N(y) \cap {\rm INV} \rrr}{\lll N(y) \rrr} \geq \frac{1}{w}$, and 
thus the probability that one iteration
fails conditioned by $y \in V_w$ is $\leq (1 - (1/w))$. Since the procedure does $n \cdot w$ iterations, the 
probability over $y \in \zon$ and over the random bits used by the algorithm $D$, conditioned by $y \in V_w$, that
$y$ is not list-inverted is $\leq (1-(1/w))^{n \cdot w} < e^{-n}.$ Therefore the probability that $y$ is
not list-inverted is bounded by the probability that $y \not \in V_w$ plus the above conditional probability of 
failure-to-list-invert. Thus,
it is bounded by $\delta - e^{-n} + e^{-n} = \delta$.

Note that the algorithm $D$ is using at each iteration the random strings
$y_1, \ldots, y_{i-1}, y_{i+1}, \ldots, y_{\ell}$ and there are $n \cdot w$ iterations. 
There is a way to fix these random strings used by $D$ so that the circuit $B$ that is obtained from $D$ by using the
fixed bits instead of random bits list-inverts a fraction of at least $(1-\delta)$ of the strings $x \in \zo^n$. There are
$n \cdot w \cdot (\ell-1)$ fixed strings.

Assuming that the circuit $C$ and $\ell$ are given, the permutation $R$ 
can be described, using the previously-discussed procedure, from 
\begin{itemize}
\item 2$\delta \cdot N \cdot n$ bits that encode the $\delta N$ elements that $B$ fails to list-invert and the value of 
$R$ at these points.
\item The last $L$ positions in the circular permutation $R$.  This requires 
$L\cdot n$ bits. 
\item For each of the $(1-\delta)N$ strings $x$ that are list-inverted by $C$, the rank of $x$ in the generated 
${\rm LIST}$.  This requires $(1-\delta)\cdot N \cdot (\log n + \log w + \log T)$ bits. 
\item
The set of $n\cdot w \cdot (\ell - 1)$ fixed strings $y$ and the value of $R$ on $y, R(y), \ldots, R^{L-1}(y)$ for
every fixed $y$. This requires  $n\cdot w \cdot (\ell - 1) \cdot n + n\cdot w \cdot (\ell - 1) \cdot L \cdot n \leq
n^2 w \ell L$ bits (for $\ell - 1 < L$). 
\end{itemize}
The total number of bits needed for the description (given $B$) is bounded by 
\[
2\delta N n + Ln + (1-\delta)N \log n + (1-\delta)N \log w + (1-\delta)N \log T +  n^2w \ell L.
\]
Plugging the values of $\ell$ and $w$, we obtain that the description of $R$ is bounded by
$2\delta N n + Ln + N \log n + (\log 6)N + N \log (1/\delta) + N \log \log (2/\gamma) + N \log (1/\gamma) 
+ N \log T  + 18n^2 \cdot L \cdot \frac{1}{\gamma} \cdot \big( \frac{1}{\delta} \big)^2 
\big( \log \frac{2}{\gamma} \big)^2$.~\qed

We want to estimate the number of permutations that are $(\gamma, T)$-good for some $L$-restricted circuit
$C$. We state the result for a particular combination of parameters that will be of interest in our application.
The extractor construction will involve the parameters $m \in \nat$ and $\epsilon > 0$. We will have $\gamma = \epsilon/m$,
$T= m^2 \cdot (1/\epsilon^2)$, and $L = m$.

\begin{lemma}
\label{l:counting} Let $n \in \nat, m \in \nat, \epsilon > 0, \delta > 0$. Let $N = 2^n$. 
Consider  $\gamma = \epsilon/m$ and 
$T = m^2 \cdot (1/\epsilon^2)$. Let $\ell = \lceil (3/\delta) \log (2/\gamma) \rceil$. Assume that $\delta = O(1)$ 
and $m^2 \cdot (1/\epsilon) = o(N/n^4).$ Let $C$ be an $m$-restricted circuit, with inputs of length $\ell n$. Then the number of
permutations $\barR$ in $\perm_\ell$ that are $(\gamma, T)$-good for $C$ is bounded by $2^h$, where $h = 3\delta\cdot N \cdot n + 3N \log m +
3 N \log (1/\epsilon)$.
\end{lemma}
$\proof$ Under the assumptions in the hypothesis, Lemma~\ref{l:description} implies that any permutation that is
$(\gamma, T)$-good for $C$ can be described with $h$ bits. The conclusion follows immediately.~\qed

\section{Analysis of the construction of pseudo-random generators from one-way permutations}

We recall the classic construction (Blum and Micali~\cite{blu-mic:j:prgen} and
Yao~\cite{yao:c:oneway}) of a pseudo-random generator from a one-way permutation. The construction starts
with a one-way permutation $\barR :\zo^{\ell n} \mapping \zo^{\ell n}$. In the classical setting, we work under
the assumption that no circuit of some bounded size inverts $\barR(\barx)$ except for a small fraction of $\barx$ in the
domain of $\barR$.
\\

\emph{Step 1.} We consider the predicate $b: \zo^{\ell n} \times \zo^{\ell n} \mapping \zo$ defined
by $b(\barx, r) = \barx \cdot r$ (the inner product modulo $2$). By the well-known Goldreich-Levin 
Theorem~\cite{gol-lev:c:one-way}, $b(\barx, r)$ is a hard-core predicate for $\barR(\barx) \odot r$, \ie, no circuit of an
appropriate bounded size can calculate $b(\barx, r)$  from $\barR(\barx) \odot r$ 
except with a probability very close to $1/2$. More precisely, it holds that if a probabilistic circuit $C_1$ on
input $\barR(\barx) \odot r$ calculates $b(\barx, r)$ with probability $1/2 + \epsilon$ (the probability is over $\barx$, $r$, and
the random bits used by $C_1$) then there is a circuit $C_2$ not much larger than $C_1$ which for a $3\epsilon/4$ fraction
of $\barx$ list-inverts $\barx$. (In the classical setting this is in conflict with the above assumption, 
because one can check the elements from the list one by one till $\barx$ is determined.) Lemma~\ref{l:step1} proves this
fact adapted to an information-theoretic context (actually, in our setting, the fact holds with stronger parameters).

\emph{Step 2.} The function $H_{\barR}: \zo^{2\ell n} \mapping \zo^{2\ell n + 1}$, given by
$H_{\barR}(\barx, r) =\barR(\barx) \odot r \odot b(\barx,r)$, can be shown to be a pseudo-random generator with extension
$1$. More precisely, it holds that if $C_2$ is a circuit that distinguishes $H_{\barR}(\barx, r)$ from $U_{2\ell n + 1}$ 
with bias $\epsilon$, one can build a circuit $C_3$, not much larger than $C_2$, that on input $\barR(\barx) \odot r$ 
calculates $b(\barx, r)$ correctly with probability at least $1/2 + \epsilon$.  Lemma~\ref{l:step2} proves this
fact adapted to an information-theoretic context.

\emph{Step 3.}  We define $G_{\barR} (\barx, r)$ by the following algorithm.
\vspace{0.75cm}

\myrule

Input: $\barR$ a permutation of $\zo^{\ell n}$, $\barx \in \zo^{\ell n}$, $r \in \zo^{\ell n}$.
\smallskip

For $i = 0$ to $m-1$, $b_i = r \cdot ( \barR^{i}(\barx))$.

Output $b_0 \odot b_1 \odot \ldots \odot b_{m-1}$.
\smallskip

\myrule

It can be shown that under the given assumption, $G_{\barR}$ is a pseudo-random generator. More precisely, it holds that
if a circuit $C_4$ distinguishes $G_{\barR}$ from $U_m$ with bias $\epsilon$, then there is a circuit $C_3$, not much
larger than $C_4$, so that $C_3$ distinguishes $H_{\barR}(\barx, r)$ from $U_{2\ell n + 1}$ with bias at least $\epsilon/m$.
 Lemma~\ref{l:step3} proves this
fact adapted to an information-theoretic  context.

We need to establish the properties of the above transformations (Steps 1, 2, and 3) in an information-theoretic
context because they will be used for the construction of an extractor. 
In our setting $\barR: \zo^{\ell n} \mapping \zo^{\ell n}$ is a random
permutation and $C_4$ is a statistical test. We will show that there are
some circuits $C_{1,1}, \ldots, C_{1,2^{m+1}-4}$ such that if $\barR$ does not hit $C_4$ $\epsilon$-correctly via $G$, then $\barR$ is $(\epsilon/m, m^2/\epsilon^2)$-good for some $C_{1,i}$, and thus, by the results in the previous section, $\barR$ has a short description.
In our context, the size of the different circuits appearing in Steps 1, 2, and 3 will be considered to be unbounded.
What matters is the number and the pattern of queries, \ie, the fact that $C_3$, $C_2$ and $C_1$ are restricted circuits. 
This is an informatic-theoretic feature.
The following lemmas follow closely the standard proofs, only that, in addition, 
they analyze the pattern of queries made by the circuits involved.

\begin{lemma} 
\label{l:step3}

\emph{(Analysis of Step 3.)} For any circuit $C_4$ there are $2^{m-1}-1$ circuits $C_{3,1}, C_{3,2}, \ldots, C_{3, 2^{m-1}-1}$ such that:
\begin{itemize}
\item[(1)] If $\barR$ is a permutation with
\[
|
\prob_{\barx, r} (G_{\barR}(\barx, r) \in C_4) - \prob( U_m \in C_4) | > \epsilon,
\]
(\ie, $\barR$ does not hit $C_4$ $\epsilon$-correctly via $G$), then there is $i \in \{1, \ldots, 2^{m-1}-1\}$ such that
\[
|\prob(H_{\barR} (U_{\ell n}, U'_{\ell n}) \in C^{\barR}_{3,i}) - \prob(U_{2\ell n +1} \in  C^{\barR}_{3,i})| > \frac{\epsilon}{m}.
\]
\item[(2)] All the circuits $C_{3,i}$ are $(m-2)$-restricted.
\end{itemize}
\end{lemma}
 $\proof$
For $k \in \{0, \ldots, m-1\}$, we define the distributions
\[
d_k = U_k \odot (U_{\ell n} \cdot U'_{\ell n}) \odot (\barR(U_{\ell n}) \cdot U'_{\ell n}) \odot \ldots \odot
(\barR^{m-k-1}(U_{\ell n}) \cdot U'_{\ell n}),
\]
where $U_k$, $U_{\ell n}$, and $U'_{\ell n}$ are distinct instances of the uniform distributions on $\zo^k$, 
$\zo^{\ell n}$, and $\zo^{\ell n}$, respectively.
Suppose that a permutation $\barR$ satisfies
\begin{equation}
\label{e:trans3}
\prob_{\barx, r}(G_{\barR}(\barx, r) \in C_4) - \prob (U_m \in C_4 ) > \epsilon.
\end{equation}
In the new notation, the above reads $\prob(d_0 \in C_4) -  \prob(d_{m-1} \in C_4) > \epsilon.$ This implies that there
is some $k \in \{0, \ldots, m-2\}$ such that $\prob(d_k \in C_4) -  \prob(d_{k+1} \in C_4) > \epsilon/m.$
For $z_1 \in \zo^{\ell n}$, $z_2 \in \zo^{\ell n}$, $z_3 \in \zo$, we define 
\[
f(z_1 \odot z_2 \odot z_3) = 
z_3 \odot (z_1 \cdot z_2) \odot (\barR(z_1) \cdot z_2) \odot \ldots \odot (\barR^{m-k-2}(z_1) \cdot z_2).
\]
Note that
\[
d_k = U_k \odot f(R(U_{\ell n}) \odot U'_{\ell n} \odot (U_{\ell n} \cdot U'_{\ell n}))
\]
and
\[
d_{k+1} = U_k \odot f(U_{\ell n} \odot U'_{\ell n} \odot U_1).
\]
We define the following 
circuit $D$ that is able to distinguish $H_{\barR}(U_{\ell n}, U'_{\ell n})$ from $U_{2\ell n +1}$. 
The input of $D$ is a string $y \in \zo^{2\ell n +1}$, which we break into $y = y_1 \odot y_2 \odot y_3$, with
$y_1$ and $y_2$ in $\zo^{\ell n}$, and $y_3 \in \zo$.
The circuit $D$ on input
$y \in \zo^{2\ell n +1}$, chooses a $k$-bits long string $u$, calculates $f(y_1, y_2, y_3)$ using the oracle $\barR$ 
and simulates $C_4$ on input $u \odot f(y_1, y_2, y_3)$.
Note that the calculation of $f(y_1, y_2, y_3)$ requires at most the query of the strings 
$y_1, R(y_1), \ldots, R^{m-3}(y_1)$. Thus, $D$ is an $(m-2)$-restricted circuit.
Clearly, $\prob_{u,y}(y \in D^{\barR}) = \prob (d_{k+1} \in C_4)$ and 
$\prob_{u,\barx, r}(H_{\barR}(\barx,r) \in D^{\barR}) = \prob (d_{k} \in C_4)$. 
Therefore, 
$\prob_{u,\barx, r}(H_{\barR}(\barx,r) \in D^{\barR}) - \prob_{u,y}(y \in D^{\barR}) > \epsilon/m.$ 
By fixing in all possible ways
$k \in \{0, \ldots, m-2\}$ and then the $k$-bits long string $u$, we obtain $2^{m-1} - 1$ circuits, denoted  
$C_{3,1}, \ldots,
C_{3, 2^{m-1}-1}$,
that act like $D$ except that
the random bits are replaced by the fixed bits. The argument above shows that if 
$R$ satisfies Equation~(\ref{e:trans3}), then
there is one circuit $C_{3,i}$ in the above set of circuits such that $\prob_{\barx, r}(H_{\barR}(\barx,r) \in C_{3,i}^
{\barR}) - 
\prob_{y}(y \in C_{3,i}^{\barR}) > \epsilon/m.$ The circuits $C_{3,i}$ are $(m-2)$-restricted circuits. 
With a similar proof, one can see that if 
\begin{equation}
\prob  (U_m \in C_4 ) - \prob_{\barx, r}(G_{\barR}(\barx, r) \in C_4) > \epsilon,
\end{equation}
then there is one circuit $C_{3,i}$ such that $\prob_{y}(y \in C_{3,i}^{\barR}) -
\prob_{\barx, r}(H_{\barR}(\barx,r) \in C_{3,i}^
{\barR}) > \epsilon/m.$~\qed

\begin{lemma} 
\label{l:step2}
\emph{(Analysis of Step 2.)} Let $C_3$ be an oracle circuit that is $L$-restricted, for some parameter $L$. 
There are four oracle circuits $C_{2,1}, C_{2,2}, C_{2,3}, C_{2,4}$ such that
\begin{itemize}
\item[(1)] If a permutation $\barR$ satisfies
\[
| \prob_{\barx, r} ( H_{\barR} (\barx \odot r) \in C_3^{\barR}) - \prob (U_{2\ell n +1} \in C_3^{\barR}) | > \epsilon,
\]
then there is $i \in \{1,2,3, 4\}$ such that
\[
\prob_{\barx, r} (C_{2,i}^R(\barR(\barx) \odot r) = b(\barx,r)) > \frac{1}{2} + \epsilon.
\]
\item[(2)] The four circuits  are $L$-restricted.
\end{itemize}
\end{lemma}
$\proof$
We define the oracle circuit $B$ that on input $\barR(\barx) \odot r$ runs as follows. It chooses a random bit $u$ and 
then it simulates the circuit $C_3^{\barR}$ to determine if  $\barR(\barx) \odot r \odot  u$ belongs to $C_3^R$ or not. 
If the answer is YES, the output is $u$, and if the answer is NO, the output is $1-u$. We also define the 
circuit $D$ in a similar way, with the only change that the YES/NO branches are permuted. Note that $B$ and $D$ are both
circuits that are $L$-restricted.

Recall that $H_{\barR}(\barx, r) = \barR(\barx) \odot r \odot b(\barx,r)$. Let us suppose that for some permutation $\barR$, 
$\prob_{\barx, r} ( \barR(\barx) \odot r \odot b(\barx,r)) \in C^{\barR}_3) - \prob (U_{2\ell n + 1} \in C_3^{\barR}) 
> \epsilon$. 
Note that $\prob_{\barx, r} ( \barR(\barx) \odot r \odot b(\barx,r) ) \in C^{\barR}_3) - 
\prob (U_{2\ell n + 1} \in C_3^{\barR}) = 
(\prob_{\barx, r} ( \barR(\barx) \odot r \odot b(\barx,r) ) \in C^{\barR}_3) - \prob_{U_1, \barx, r} 
(  \barR(\barx) \odot r \odot U_1 ) \in C_3^{\barR})) +
(\prob_{U_1, \barx, r} (  \barR(\barx) \odot r \odot U_1) \in C_3^{\barR}) - \prob (U_{2\ell n + 1} \in C_3^{\barR}))$. 
The second term is equal to zero, because $R$ is a permutation and, thus, $ U_1 \odot \barR(\barx) \odot r $ 
is actually the uniform distribution on $\zo^{2\ell n+1}$. Therefore, 
$\prob_{\barx, r} ( \barR(\barx) \odot r \odot b(\barx,r)) \in C^{\barR}_3) - 
\prob_{U_1, \barx, r} ( \barR(\barx) \odot r \odot U_1) \in C_3^{\barR}) > \epsilon$. 
According to Yao's lemma that connects predictors to distinguishers 
(for a proof see, for example,~\cite[pp. 162]{zim:b:book}), it follows 
that $\prob_{u, \barx, r}(B^R(\barR(\barx) \odot r) = b(\barx,r)) > \frac{1}{2} + \epsilon$. 
Let $B_0$ ($B_1$) 
be the circuit that is obtained from $B$ by fixing bit $u$ to $0$ (respectively, to $1$). 
Then at least one of the events ``$B_0 (\barR(\barx) \odot R) = b(\barx, r)$" or ``$B_1 (\barR(\barx) \odot R) = 
b(\barx, r)$" has probability $> \frac{1}{2} + \epsilon.$ 

If $\prob_{z} (z \in C_3^{\barR}) - \prob_{\barx, r} ( \barR(\barx) \odot r \odot b(\barx,r)) \in C^{\barR}_3) > \epsilon$, then 
the same argument works for the circuit $D$, and we obtain two deterministic circuits $D_0$ and $D_1$. The four circuits
$B_0, B_1, D_0$ and $D_1$ satisfy the requirements.~\qed

\begin{lemma} 
\label{l:step1}
\emph{(Analysis of Step 1.)}
Let $C_2$ be an oracle circuit that is $L$-restricted for some parameter $L$. Then there is a circuit $C_1$ such that
\begin{itemize}
\item[(1)] If $\barR$ is a permutation such that $\prob_{\barx, r} (C_2^{\barR}(\barR(\barx) \odot r) = b(\barx, r)) 
> \frac{1}{2} + \epsilon$, then for at least a fraction $\epsilon$ of $\barx \in \zo^{\ell n}$, $C_1^{\barR}$ on input
$\bary = \barR(\barx)$ outputs a list of $1/\epsilon^2$ strings that contains $\barx$ (\ie, $\barR$ is 
$(\epsilon, 1/\epsilon^2)$-good for $C_1$).
\item[(2)]
The circuit $C_1$ is $L$-restricted.
\end{itemize}
\end{lemma}
$\proof$
Suppose permutation $\barR: \zon \mapping \zo^{\ell n}$ satisfies 
$\prob_{\barx, r}(C_2^{\barR} (\barR(\barx) \odot r) = b(\barx, r)) > (1/2) + \epsilon$. 
Then, by a standard averaging argument, for a fraction $\epsilon$ of $\barx$ in $\zo^{\ell n}$, 
$\prob_{r}(C_2^R (\barR(\barx) \odot r) = b(\barx, r)) > (1/2) + (\epsilon/2)$.  Consider such an $\barx$ and let
${\rm Had}(\barx)$ denote the encoding of $\barx$ via the Hadamard error-correcting code (see~\cite{tre:t:codesurvey}).
By the definition of the Hadamard code, $b(\barx, r)$ is just the $r$-th bit of ${\rm Had}(\barx)$.
Thus the string
$u = C_2^R(\barR(\barx) \odot (0 \ldots 0)) \odot \ldots \odot C_2^R(\barR(\barx) \odot (1 \ldots 1))$ agrees with 
${\rm Had}(\barx)$ on at least a fraction $(1/2) + (\epsilon/2)$ of positions. Since the circuit 
$C_2$ is $L$-restricted, the string $u$ can be calculated by querying only $\bary$, 
$\barR(\bary), \ldots, \barR^{L-1}(\bary)$, where $\bary = \barR(\barx)$. By brute force we can determine the
list of all strings $\barz$ so that ${\rm Had}(\barz)$ agrees with $u$ in at 
least $\frac{1}{2} + \frac{\epsilon}{2}$ positions. It is known (see, for example, ~\cite[pp. 218]{zim:b:book})
that
there are at most $\frac{1}{4} \cdot \big( \frac{2}{\epsilon}\big)^2 =
\big( \frac{1}{\epsilon}\big)^2$ such strings $\barz$ and one of them is $\barx$.~\qed

By combining Lemma~\ref{l:step3}, Lemma~\ref{l:step2}, and Lemma~\ref{l:step1}, we obtain the following fact. 

\begin{lemma}
\label{l:analysis}
Let $C_4$ be a circuit. Then there are $2^{m+1}-4$ circuits $C_{1,1}, \ldots, C_{1,2^{m+1}-4}$ such that
\begin{itemize}
\item[(1)] If $\barR$ is a permutation with
\[
|
\prob_{\barx, r} (G_{\barR}(\barx, r) \in C_4) - \prob( U_m \in C_4) | > \epsilon,
\]
(\ie, $\barR$ does not hit $C_4$ $\epsilon$-correctly via $G$), then there is some circuit
$C_{1,i}$ such that for at least a fraction $\frac{\epsilon}{m}$ of $\barx$, $C_{1,i}^{\barR}$ on input 
$\barR(\barx)$ outputs a list of
$m^2 \cdot \big(\frac{1}{\epsilon}\big)^2$ strings that contains $\barx$ (\ie, $\barR$ is $(\epsilon/m, m^2/\epsilon^2)$-good
for $C_{1,i}$).
\item[(2)] All the circuits $C_{1,i}$ are $(m-2)$-restricted.

\end{itemize}
\end{lemma}

\section{An extractor from a crypto pseudo-random generator}
\label{s:first}

We first build a special type of extractor in which the weakly-random string is the truth-table of a permutation in
$\perm$.

The following parameters will be used throughout this section. 
Let $\epsilon >0, \delta > 0$, and $n, m \in \nat$ be parameters. Let $N = 2^n$.
Let $\ell = \lceil (3/\delta) \log (2 m \cdot (1/\epsilon)) \rceil$. We
consider the set of permutations $\perm_\ell$. We assume that $\delta = O(1)$ and $m^2 \cdot (1/\epsilon) = o(N/n^4).$

Let $G : \perm_\ell \times (\zo^{\ell n} \times \zo^{\ell n}) \mapping \zo^m$ be the function defined by
the following algorithm (the same as the algorithm for $G_{\barR}$ from the previous section).
\medskip

\myrule

Parameters: $\ell \in \nat, m \in \nat$.

Input: $\barR \in \perm_\ell$, $(\barx, r) \in \zo^{\ell n} \times \zo^{\ell n}$.

For $i = 0$ to $m-1$, $b_i = r \cdot \barR^i(\barx)$.

Output $b_0 \odot b_1 \odot \ldots \odot b_{m-1}$.

\myrule

The following lemma, in view of Lemma~\ref{l:extractor1}, shows that $G$ is an extractor for the special case of
weakly-random strings that are truth-tables of permutations in $\perm_\ell$.

\begin{lemma}
\label{l:badhitters}
Let $C_4$ be a test for strings of length $m$ (\ie, $C_4 \subseteq \zo^m$). Let
${\rm GOOD}(C_4) = \{ \barR \in \perm_\ell \mid \mbox{$\barR$ does not hit $C_4$ $\epsilon$-correctly via $G$}\}$.
Then $\lll {\rm GOOD}(C_4) \rrr < 2^{m+h+1}$, where $h = 3\delta N n + 3N \log m + 3N \log (1/\epsilon)$.
\end{lemma}
$\proof$ Let $C_{1,1}, \ldots, C_{1, 2^{m+1}-4}$ be the $2^{m+1}-4$ circuits implied by
Lemma~\ref{l:analysis} to exist (corresponding to the test $C_4$). Let $\barR$ be in ${\rm GOOD}(C_4)$. Then 
Lemma~\ref{l:analysis} shows that there is a circuit $C_{1,i}$ from the above list having the following property:
For at least a fraction $\gamma = \epsilon/m$ of strings $\barx \in \zo^{\ell n}$, $C_{1,i}^{\barR}$ on input $\barR(\barx)$
returns a list having $T = m^2 \cdot (1/\epsilon^2)$ strings, one of which is $\barx$. Thus, $\barR$ is
$(\gamma, T)$-good for $C_{1,i}$ (recall Definition~\ref{d:goodperm}). It follows that the set of permutations
$\barR \in \perm_\ell$ that do not hit $C_4$ $\epsilon$-correctly via $G$ is included in 
$\bigcup_1^{2^{m+1}-4} \{ \barR \in \perm_\ell \mid \mbox{$\barR$ is $(\gamma, T)$-good for $C_{1,i}$} \}$.
Lemma~\ref{l:counting} shows, that, for each $i \in \{1, \ldots, 2^{m+1}-4 \}$, 
$\lll \{ \barR \in \perm_\ell \mid \mbox{$\barR$ is $(\gamma, T)$-good for $C_{1,i}$} \} \rrr \leq 2^h$, where
$h = 3\delta\cdot N \cdot n + 3N \log m +
3 N \log (1/\epsilon)$. The conclusion follows.~\qed
\medskip

In order to obtain a standard extractor (rather than the special type given by Lemma~\ref{l:badhitters}), the only
thing that remains to be done is to transform a random binary string $X$ into a permutation $R \in \crc$, which
determines $\barR \in \perm_{\ell}$ that is used in the function $G$ given above.

Note that a permutation $R \in \crc$ is specified by $(R(1), R^2(1), \ldots, R^{N-1}(1))$, which is an arbitrary 
permutation of the set $\{2, 3, \ldots, N\}$. Consequently, we need to generate permutations of the set
$\{1, 2, \ldots, N-1\}$ (which can be viewed as permutations of $\{2, 3, \ldots, N\}$ in the obvious way). We can use
the standard procedure that transforms a function mapping $[N-1]$ to $[N-1]$ 
into a permutation of the same type. To avoid some minor truncation nuisances, we actually use a function $X: [N] 
\mapping [N]$.

\myrule

Input: $X: [N] \mapping [N]$.

\quad \quad for $i = 1$ to $N-1$, $R(i)=i$ (initially $R$ is the identity permutation).

\quad \quad {\bf Loop 2:}

\quad \quad for $i = 1$ to $N-1$

\quad \quad \quad \quad $Y(i) = 1 + (X(i) \mod i)$.

\quad \quad {\bf Loop 3:}

\quad \quad for $i = 1$ to $N-1$

\quad \quad \quad \quad Swap $R(i)$ with $R(Y(i))$.

Output: permutation $R: [N-1] \mapping [N-1]$.

\myrule

We want to estimate the number of functions $X: [N] \mapping [N]$ that map via the above procedure to a given
permutation $R: [N-1] \mapping [N-1]$. We call a sequence $(Y(1), \ldots, Y(N))$ a \emph{*-sequence} if, for all $i$, 
$Y(i) \in \{1, \ldots, i\}$. 
Observe that, using Loop 3, a *-sequence $(Y(1), \ldots, Y(N-1))$ defines a unique permutation $(R(1), \ldots, R(N-1))$,
and thus it is enough to estimate the maximum number of functions $X: [N] \mapping [N]$ that map via Loop 2 in the 
above procedure in a given *-sequence $(Y(1), \ldots, Y(N-1))$ (the maximum is taken over all *-sequences
of length $N-1$). We denote this number by $A(N)$. A (rough) upper bound can be established as follows. 
\begin{eqnarray*}
A(N) & \leq & \Big \lceil \frac{N}{1} \Big \rceil \cdot  \Big \lceil \frac{N}{2} \Big \rceil \cdot  \ldots 
 \cdot \Big \lceil \frac{N}{N-1} \Big \rceil \\
& \leq & \Big( \frac{N}{1} + 1 \Big) \cdot \Big( \frac{N}{2} + 1 \Big) \cdot \ldots \cdot \Big( \frac{N}{N-1} + 1 \Big) \\
& = & \frac{(N+1) (N+2) \cdot \ldots \cdot (2N-1)}{1 \cdot 2 \cdot \ldots \cdot (N-1)} \\
&  = & {2N-1 \choose N - 1}  \leq 2^{2N}.
\end{eqnarray*}

We can now present the (standard) extractor. 
We choose the parameters as follows. Fix $n \in \nat$ and let $N = 2^n$ and $\barN = n \cdot 2^n$. 
Let $\lambda \in (0, 1)$ be a constant. Let $\alpha > 0, \beta > 0$ be constants such
that $\alpha < \lambda/3, \beta < (\lambda - 3\alpha)/4$. Let $\epsilon \geq N^{-\beta}$ and $m \leq N^\alpha$.
Take $\delta = (\lambda - 4\beta - 3\alpha)/4$
and $\ell = \lceil (3/\delta) \log (2 m \cdot (1/\epsilon))\rceil$.
The weakly-random string $X$ has length $\barN$ and is viewed as the truth-table of
a function mapping $[N]$ to $[N]$. The seed is of the form $y = (\barx, r) \in \zo^{\ell n} \times \zo^{\ell n}$.
We first transform $X$ into a permutation $\barR(X) \in \perm_{\ell}$ using the above algorithm and then taking the
$\ell$-product. We define the extractor $E: \zo^{\barN} \times \zo^{2\ell n} \mapping \zo^m$
by $E(X, (\barx , r)) = G(\barR(X), (\barx , r))$. More explicitly, the extractor is defined
by the following procedure.

\myrule

Parameters: $n \in \nat, \barN \in \nat$, $\lambda > 0, \epsilon > 0$, $\ell \in \nat, m \in \nat$, satisfying the above
requirements.
\\

Inputs: The weakly-random string $X \in \{0,1\}^{\barN}$, viewed as the truth-table of a function $X: [N] \mapping [N]$; the seed $y = (\barx, r) \in \zo^{\ell n} \times  \zo^{\ell n}$.
\\

Step 1. Transform $X$ into a permutation $\barR_X \in \perm_{\ell}$. The transformation is performed by the above procedure
which yields a permutation $R \in \crc$, and, next, $\barR_X$ is the $\ell$-direct product of $R$.
\\

Step 2. For $i = 0$ to $m-1$, $b_i = r \cdot \barR^{i}_X(\barx)$.

Output $b_0 \odot b_1 \odot \ldots \odot b_{m-1}$, which is denoted $E(X, y)$.
\\

\myrule

We have defined a function $E: \zo^{\barN} \times \zo^{2\ell n} \mapping \zo^m$. Note that the seed length
$2\ell n$  is $O(\log^2 \barN)$ and the output length $m$ is $\barN^{\alpha}$, for an arbitrary $\alpha < \lambda/3$.

\begin{theorem}
The function $E$ is a $(\lambda \barN, 2\epsilon)$-extractor.
\end{theorem}
$\proof$ 
Let $C_4$ be a subset of $\zo^m$. 
Taking into account Lemma~\ref{l:extractor1}, it is enough to show that the number of strings $X \in \zo^{\barN}$ that do
not hit $C_4$ $\epsilon$-correctly via $E$ is at most $2^{\lambda \barN - \log(1/\epsilon)}$. Let
 $X \in \zo^{\barN}$ be a string that does not hit $C_4$ $\epsilon$-correctly via $E$. By the definition of $E$, it
follows that $\barR_X$ does not hit $C_4$ $\epsilon$-correctly via $G$. By Lemma~\ref{l:badhitters}, there are at most
$2^{m+h+1}$ permutations $\barR \in \perm_{\ell}$ that do not hit $C_4$ $\epsilon$-correctly via $G$, where $h = 3\delta N n
+ 3N \log m + 3N \log(1/\epsilon)$. Since the number of functions $X : [N] \mapping [N]$ that map into a given
permutation $\barR \in\perm_{\ell}$ is at most $A(N) < 2^{2N}$, it follows that
\[
\lll \{ X \in \zo^{\barN} \mid \mbox{$X$ does not hit $C_4$ $\epsilon$-correctly}\} \rrr < 2^{2N} \cdot 2^{m+h+1} < 
2^{\lambda \barN - \log(1/\epsilon)},
\]
where the last inequality follows from the choice of parameters.~\qed

\section{A bitwise locally-computable extractor}
\label{s:secondext}

We present a bitwise locally-computable extractor: 
Each bit of the output string can be calculated separately in $O(\log^2 \barN)$, where $\barN$ is
the length of the weakly-random string. The proof uses the same plan as for the extractor in Section~\ref{s:first}, except that the weakly-random string $X$ is viewed as the truth-table of an arbitrary function (not necessarily a permutation) and the ``consecutive"' values that are used in the extractor are $\barX(\barx), \barX(\barx+1), \ldots, \barX(\barx + m-1)$ (instead of $\barR(\barx), 
\barR^2(\barx), \ldots, \barR^{m-1}(\barx)$ used in  Section~\ref{s:first}).

The parameter $n \in \nat$  will be considered fixed throughout this section.  We denote $N = 2^n$ and $\barN = n \cdot N$.
The parameter $m \in \nat$ will be specified
later (it will be a subunitary power of $N$). 
For two binary strings $x$ and $r$ of 
the same length, $b(x,r)$ denotes the inner product of $x$ and $r$ viewed as vectors over the field ${\rm GF}(2)$.

The weakly-random string $X$ has length $\barN$, and is viewed as the truth-table of
a function $X: \zon \mapping \zon$. For some $\ell \in \nat$ that will be specified later we define
$\barX: \zo^{\ell n} \mapping \zo^{\ell n}$ by $\barX (x_1 \odot \ldots \odot x_{\ell}) = X(x_1) \odot \ldots 
\odot X(x_\ell)$, \ie, $\barX$ is the $\ell$-direct product of $X$. We also denote $\barx = x_1 \odot \ldots \odot x_\ell$. The seed of the extractor will be 
$(\barx, r) \in \zo^{\ell n} \times \zo^{\ell n}$. We define $\barx + 1 = (x_1 + 1) \odot \ldots \odot (x_\ell + 1)$ (where 
the addition is done modulo $2^n$) and inductively, for any $k \in \nat$, $\barx + k + 1 = (\barx + k) + 1$. The extractor 
is defined by
\begin{equation}
\label{e:localextractor}
E(X, (\barx, r)) = b(\barX(\barx), r) \odot  b(\barX(\barx+1), r) \odot \ldots \odot  b(\barX(\barx+m-1), r).
\end{equation}
A set $D \subseteq \zo^m$ is called a \emph{test}. We say that $X$ hits a test $D$ $\epsilon$-correctly via $E$ if
$|\prob_{\barx, r} (E(X, (\barx, r)) \in D) - \frac{\lll D \rrr}{\lll \zo^m \rrr} | \leq \epsilon$.
We want to show that the number of functions $X$ that do not hit $D$ $\epsilon$-correctly via $E$ is small and then
use Lemma~\ref{l:extractor1}. To this aim we investigate the properties 
of a function $X$ that does not hit a test $D \subseteq \zo^m$ $\epsilon$-correctly
via $E$.
\begin{lemma}
\label{l:predictors}
Let $D \subseteq \zo^m$ be a fixed set. Then there are $2^{m+2}-4$ circuits $C_1, \ldots C_{2^{m+2}-4}$ such that
if $X$ does not hit $D$ $\epsilon$-correctly, then there is some circuit $C_i$, $i \in \{1, \ldots, 2^{m+2}-4\}$, 
such that
\[
\prob_{\barx, r} (C_i \mbox{ on input  
$b(\barX(\barx-m+1), r) \odot  \ldots \odot  b(\barX(\barx-1), r)$ outputs $b(\barX(\barx), r)$}) \geq  1/2 + \epsilon/m.
\]
\end{lemma}
$\proof$
The proof is similar to the proofs of lemmas~\ref{l:step3} and~\ref{l:step2}. Let $X: \zo^{n} \mapping \zo^{n}$ be a function that does not hit $D$ $\epsilon$-correctly via $E$. This means 
that $| \prob_{\barx, r}(E(X, (\barx,r)) \in D) - \prob(U_m \in D)| > \epsilon$. Let us first suppose that
\begin{equation}
\label{e:hybridglobal}
\prob_{\barx, r}(E(X, (\barx, r)) \in D) - \prob(U_m \in D) > \epsilon.
\end{equation}
For each $k \in \{0, \ldots, m-1\}$, we define the hybrid distribution $d_k$ given by
\[
d_k = b(\barX(\barx),r) \odot b(\barX(\barx+1),r) \odot \ldots \odot b(\barX(\barx+m-k-1),r) \odot U_{k}.
\]
and
\[ 
d_m = U_m.
\]
Equation~(\ref{e:hybridglobal}) states that $\prob (d_0 \in D) - \prob (d_{m} \in D) > \epsilon$. Using the
standard argument, it follows that there exists $k \in \{0, \ldots, m-1\}$ such that $\prob (d_k \in D) 
- \prob (d_{k+1} \in D) > \epsilon/m$. We build a probabilistic circuit $C$ that on input
$b(\barX(\barx),r) \odot b(\barX(\barx + 1),r) \odot \ldots \odot b(\barX(\barx + m -k-2),r)$ attempts to calculate 
$b(\barX(\barx + m - k - 1),r)$.

\myrule

Circuit $C$.

Input: $v_0 \odot v_1 \odot \ldots \odot v_{m-k-2}$, each $v_i \in \zo$. (In case $k=m-1$, there is no input.)
\\

Choose randomly $u \in \zo, t \in \zo^k$.
\\

If $v_0 \odot v_1 \odot \ldots \odot v_{m-k-2} \odot u \odot t \in D$, return $u$.

Else return $1-u$.

\myrule

By Yao's lemma on predictors versus distinguishers, it holds that
\[
\prob_{\barx, r} (C \mbox{ on input  
$b(\barX(\barx), r) \odot  \ldots \odot  b(\barX(\barx+m-k-2), r)$ outputs $b(\barX(\barx+m-k-1), r) $}) 
\geq  1/2 + \epsilon/m,
\]
where the probability is taken over $\barx \in \zo^{\ell n}, r \in \zo^{\ell n}$ and the random bits used by $C$. The procedure $C$
uses $k+1$ random bits (for $u$ and $t$), and $k \in \{0, \ldots, m-1\}$. By considering all possibilities for $k$
and by fixing the $(k+1)$ bits in all possible ways, we obtain $2^1 + \ldots + 2^m = 2^{m+1}-2$ circuits
$C_1, \ldots, C_{2^{m+1}-2}$ with the desired property. 

In the alternative (to Equation~(\ref{e:hybridglobal})) case
\[
\prob(U_m \in D) - \prob(E(X,(\barx,r)) \in D) > \epsilon,
\]
we obtain in a similar way another set of $2^{m+1}-2$ circuits.~\qed

The next lemma is an analogue of Lemma~\ref{l:step1}. It states that if there exists a circuit $C$ with the property
indicated in Lemma~\ref{l:predictors}, then, from $\barX(\barx - m +1), \ldots, \barX(\barx-1)$, one can compute,
in a weak but non-trivial way, $\barX(\barx)$.
\begin{lemma}
\label{l:step1rev}
Let $C$ be a circuit. Then there is a circuit $B$ such that the following holds. Suppose $\barX : \zo^{\ell n} \mapping
\zo^{\ell n}$ is a function such that
\[
\prob_{\barx, r} (C(b(\barX(\barx - m +1), r) \odot \ldots \odot b(\barX(\barx - 1), r)) = b(\barX(\barx, r))) \geq (1/2) + 
(\epsilon/m).
\]
Then, for at least a fraction $\epsilon/m$ of $\barx$  in $\zo^{\ell n}$, $B$ on input $\barX(\barx - m +1) \odot \ldots
\odot \barX(\barx - 1)$ outputs a list of $m^2 \cdot (1/\epsilon^2)$ elements, one of which is $\barX(\barx)$.
\end{lemma}
$\proof$ The proof is similar to the proof of Lemma~\ref{l:step1}. Let $C$ and $\barX$ be as in the hypothesis. By an
averaging argument, it follows that for a fraction $(\epsilon/m)$ of $\barx$ in $\zo^{\ell n}$,
\begin{equation}
\label{e:predict}
\prob_r (C(b(\barX(\barx - m +1), r) \odot \ldots \odot b(\barX(\barx - 1), r) = b(\barX(\barx, r))) \geq (1/2) + 
\epsilon/(2m).
\end{equation}
Let ${\rm Had}( x)$ denote the encoding of a string $x$ via the Hadamard error-correcting code. 
By the definition of the Hadamard code, $b(x, r)$ is just the $r$-th bit of ${\rm Had}(x)$. Consider
the binary string $u(\barx) \in \zo^{2^{\ell n}}$ whose $r$-th bit is $C(b(\barX(\barx - m +1), r) \odot \ldots 
\odot b(\barX(\barx - 1), r)$ (here $r \in \{0, \ldots, 2^{\ell n}-1\}$ is written in base $2$ on $\ell n$ bits for the sake
of the definition of $b$). 
Clearly, the string $u(\barx)$ can be calculated from $\barX(\barx - m + 1), \ldots, 
\barX(\barx-1)$. The equation~(\ref{e:predict}) implies that, for a fraction $(\epsilon/m)$ of $\barx \in \zo^{\ell n}$,
$u(\barx)$ agrees with ${\rm Had}(\barX(\barx))$ on at least $1/2 + \epsilon/(2m)$ positions. 
By brute force, we can determine 
all the strings $z$ so that ${\rm Had}(z)$ agrees with $u(\barx)$ in at 
least $\frac{1}{2} + \epsilon/(2m)$ positions.
It is known 
that
there are at most $\frac{1}{4} \cdot \big( \frac{2m}{\epsilon}\big)^2 =
\big( \frac{m}{\epsilon}\big)^2$ such strings $z$ and, by the above discussion,  one of them is $\barX(\barx)$.~\qed

The key property of the circuit $B$ in the above lemma is captured in the following definition (which is analogous to
Definition~\ref{d:goodperm}).
\begin{definition}
\label{d:goodfunction}
Let $B$ be a circuit. A function $\barX: \zo^{\ell n} \mapping \zo^{\ell n}$ is $(\gamma, T)$-good for $B$ if for
at least a $\gamma$ fraction of $\barx \in \zo^{\ell n}$, $B$ on input $\barX(\barx - m +1) \odot \ldots
\odot \barX(\barx - 1)$ outputs a $T$- list of strings, one of which is $\barX(\barx)$.
\end{definition}
We choose the parameters in the same way as in Section~\ref{s:randperm}. The parameters $\epsilon$ and $m$ will be specified
later. We take $\delta > 0$, $\gamma = \epsilon/m$, 
$T = m^2/\epsilon^2$,
$\ell =\lceil (3/\delta) \log (2/\gamma) \rceil$ and $w = \lceil 6 \cdot (1/\delta) \cdot \log(2/\gamma) 
\cdot (1/\gamma) \rceil$.  

The next two lemmas are the analogues of Lemma~\ref{l:description}. The first lemma shows the amplification effect obtained by taking the $\ell$-direct product.
\begin{lemma} 
\label{l:amplification}
The parameters are as specified above. Let $B$ be a circuit. Then there is an oracle circuit $A$ such that:

(1) If $\barX$ is $(\gamma, T)$-good for $B$, then, for a fraction $(1-\delta)$ of $x$ in $\zon$, the circuit $A$, on input
$x$ and $X(x-m+1) \odot \ldots \odot X(x-1)$ and with access to oracle $X$ restricted as shown in (2), outputs a list 
containing $n \cdot w \cdot T$ elements, one of which is $X(x)$.

(2) The oracle circuit $A$ queries a set of $n \cdot w \cdot (\ell-1) \cdot (m-1)$ strings that do not depend on the input.
\end{lemma}
$\proof$ The proof is very similar to the first part of the proof of Lemma~\ref{l:description}. Let ${\rm GOOD}$ be the set of strings $\barx$ 
in $\zo^{\ell n}$ such that the circuit $B$, on input $\barX(\barx - m+1) \odot \ldots \odot \barX(\barx - 1)$, calculates
a $T$-list that contains $\barX(\barx)$. By hypothesis, $\lll {\rm GOOD} \rrr \geq \gamma \cdot 2^{\ell n}$. We consider the
following algorithm $A'$ that can query the oracle $X: \zon \mapping \zon$ in several random positions. 

\myrule

Input: $x \in \zon$, and $X(x-m+1), \ldots, X(x-1) \in (\zon)^{m-1}$. The algorithm can pose random queries to the
oracle $X: \zon \mapping \zon$. The goal is to calculate a list of strings that contains $X(x)$. 

${\rm LIST} = \emptyset.$

Repeat the following $n \cdot w$ times.

Pick random $i \in \{1, \ldots, \ell\}$.

Pick $\ell-1$ random strings in $\zon$ denoted $x_{1}, \ldots, x_{i-1}, x_{i+1}, \ldots, x_\ell$.

By querying the oracle $X$, find, for each $x_j$, the strings 
$X(x_j - m +1), X(x_j - m + 2), \ldots, X(x_j - 1)$. 

Let $\barx = (x_1, \ldots, x_{i-1}, x, x_{i+1}, \ldots, x_\ell)$. Build the string $\barX(\barx-m+1) \odot \barX(\barx-m+2) 
\odot \ldots \odot \barX(\barx-1)$. Run the circuit $B$ on input $\barX(\barx-m+1) \odot \barX(\barx-m+2) 
\odot \ldots \odot \barX(\barx-1)$. 

The circuit $B$ returns a $T$-list of $\ell$-tuples in $(\zon)^\ell$.

(Note: In case of success, one of these $\ell$-tuples is 

$\barX(\barx) = X(x_1), \ldots, X(x_{i-1}), X(x), X(x_{i+1}) \ldots
,  X(x_{\ell})$)

Add to ${\rm LIST}$ the $i$-th component of every $\ell$-tuple in the list produced by $B$.

End Repeat

\myrule

We say that the above algorithm is successful on input $x$ if, at the conclusion of the algorithm, 
${\rm LIST}$ contains $X(x)$. We estimate the success probability of the above circuit on input $x$. 
Let $N(x)$ be the multiset of $\ell$-tuples having $x$ as one component where the multiplicity of a tuple is the number of
occurrences of $x$ in the tuple. On input $x$, at each iteration, the algorithm chooses
uniformly at random $\overline{x}$ in $N(x)$. The algorithm succeeds at that iteration if and only if 
$\barx \in {\rm GOOD}$. By following the same arguments and the same calculations as in Lemma~\ref{l:description}, 
we conclude that the probability that algorithm $A'$ succeeds on $x$ is at least $(1-\delta)$, where the
probability is taken over $x$ and the random strings used by $A'$. Note that the algorithm
$A'$ is using at each iteration the random strings $x_1, \ldots, x_{i-1}, x_{i+1}, \ldots, x_\ell$, and there are
$n \cdot w$ iterations. For each such random string $x_j$, $A'$ needs the $(m-1)$ values $X(x_j-m+1), X(x_j-m+2), \ldots, 
X(x_j-1)$. There is a way to fix the above  random strings so that the circuit $A$, which results from $A'$ by using the fixed 
strings  instead of the random strings, succeeds on at least a $(1-\delta)$ fraction of the strings $x \in \zon$. Therefore,
the circuit $A$ has the desired properties.~\qed

\begin{lemma} 
\label{l:descriptionfunc}
The parameters are as specified above. Let $A$ be an oracle circuit and $X: \zon \mapping \zon$ be a function such that $A$ and $X$ satisfy the
conditions (1) and (2) in Lemma~\ref{l:amplification}. More precisely, we assume that:

(1) For a fraction $(1-\delta)$ of $x$ in $\zon$, the circuit $A$, on input $x$ and 
$X(x-m+1) \odot \ldots \odot X(x-1)$ and with access to oracle $\barX$ restricted as shown in (2), outputs a list 
containing $n \cdot w \cdot T$ elements, one of which is $X(x)$.

(2) The oracle circuit $A$ queries a set of $n \cdot w \cdot (\ell-1) \cdot (m-1)$ strings that do not depend on the input. 

Then, given $A$, $X$ can be described using a number of bits bounded by 
$2\delta N n + mn + N \log n + (\log 6)N + N \log (1/\delta) + N \log \log (2/\gamma) + N \log (1/\gamma) 
+ N \log T  + 36n^2 \cdot m \cdot \frac{1}{\gamma} \cdot \big( \frac{1}{\delta} \big)^2 
\big( \log \frac{2}{\gamma} \big)^2$. 
\end{lemma}
$\proof$  The oracle circuit $A$ allows a short description of the strings $X(x)$ for the fraction of $(1-\delta)$ of
the strings $x \in \zon$ given in assumption~(1). Namely, such a string $X(x)$ is completely determined by the circuit $A$,  by the value
of $X$ for the fixed set of queries given in assumption~(2), by the
previous $m-1$ values $X(x-m+1), \ldots, X(x-1)$, and by the rank of $X(x)$ in the list returned by $A$ on input $x,
X(x-m+1), \ldots, X(x-1)$. Thus, the truth-table of the function $X: \zon \mapping \zon$ can be described (given the circuit $A$) using
the following information.
\begin{itemize}
\item 2$\delta \cdot N \cdot n$ bits that encode the set of $\delta N$ elements on which $A$ fails  
and the value of $X$ at these points.
\item The ``first'' $m-1$ values $X(0), \ldots, X(m-1)$.  This information requires 
$(m-1) \cdot n$ bits. (Here, $X(i)$ represents the value of $X$ at the $i$-th string in $\zon$, lexicographically ordered.)
\item For each of the $(1-\delta)N$ strings $y$ on which $A$ succeeds, the rank of $X(x)$ in the list returned 
by $A$.  This requires $(1-\delta)\cdot N \cdot (\log n + \log w + \log T)$ bits.  
\item
The set of $n\cdot w \cdot (\ell - 1) \cdot (m-1)$ fixed strings that are queried by $A$ and the value of $X$ at these
strings.  This information requires  $2n^2\cdot w \cdot (\ell - 1) \cdot (m-1)$ bits.
\end{itemize}
The total number of bits needed for the description of $X$ (given $A$) is bounded by
\[
2\delta N n + m\cdot n + (1-\delta)N \log n + (1-\delta)N \log w + (1-\delta) N \log T + 
2n^2\cdot w \cdot (\ell - 1) \cdot (m-1).
\]
Keeping into account that $\ell =\lceil (3/\delta) \log (2/\gamma) \rceil$ 
and $w = \lceil 6 \cdot (1/\delta) \cdot \log(2/\gamma) 
\cdot (1/\gamma) \rceil$, the conclusion follows.~\qed 

We make the final choice of parameters. Let $n \in \nat$ and the constant $\lambda \in (0,1)$. Recall that $N = 2^n$ and
$\barN = n\cdot N$. We take the constants $\alpha < \lambda/3, \beta < (\lambda - 3\alpha)/4$ and $\delta = (\lambda -3\alpha - 4\beta)/4$.
We also take the output length $m \leq N^\alpha$ and the extractor bias $\epsilon \geq N^{-\beta}$. Note that
$\ell  = \lceil (3/\delta) \log (2m/\epsilon) \rceil = O(n)$.
\begin{theorem}
\label{t:localextractor} Assume that the parameters $\barN, \lambda, m, \ell$ and $\epsilon$ satisfy the above
requirements. Then the function $E: \zo^{\barN} \times \zo^{2\ell n} \mapping \zo^m$, given in 
Equation~\ref{e:localextractor}, is a $(\lambda \barN, 2\epsilon)$-extractor.
\end{theorem}
$\proof$
Assume $X \in \zo^{\barN}$ does not hit a test $D \subseteq \zo^m$ $\epsilon$-correctly via $E$. Then $X$ can be
described by one of the circuits $C_1, \ldots, C_{2^{m+2}-4}$, given by Lemma~\ref{l:predictors}, and, according
to Lemma~\ref{l:descriptionfunc}, by a string of length $h$, where $h \leq 2\delta N n + mn + N \log n + (\log 6)N + N \log (1/\delta) + N \log \log (2/\gamma) + N \log (1/\gamma) 
+ N \log T  + 36n^2 \cdot m \cdot \frac{1}{\gamma} \cdot \big( \frac{1}{\delta} \big)^2 
\big( \log \frac{2}{\gamma} \big)^2$. Thus, the number of strings $X$ that do not hit $D$ $\epsilon$-correctly via $E$ is bounded
by $2^{m+2+h}$. For our choice of parameters, it holds that $m+2+h \leq \lambda \barN - \log(1/\epsilon)$. Therefore,
by Lemma~\ref{l:extractor1}, $E$ is a $(\lambda \barN, 2 \epsilon)$-extractor.~\qed

The construction scheme of the last extractor (given in Equation~(\ref{e:localextractor})) allows some flexibility in
the choice of parameters and, in particular, we can obtain an extractor with seed length logarithmic in the
length of the weakly random string. Namely, we can consider the weakly random string $X$ to be the truth-table of a
function of type $X: \zon \mapping \zo^{N_1}$, where $N_1 >> n$. We use the same value of $\ell$, and we take the $\ell$-direct
product of $X$ and obtain $\barX: \zo^{\ell n} \mapping \zo^{\ell N_1}$. Clearly, $|\barX(\barx)| = \ell N_1$.
To get a short seed we need to replace the Hadamard code (recall that the function $b(x,r)$ gives the $r$-th bit of
${\rm Had}(x)$) by an error-correcting code with a good list decoding property that has a better rate. For example
the code given in~\cite{ghsz:j:listcoding}, which we denote ${\rm Code}$, is of the type ${\rm Code}: \zo^{\tn} \mapping \zo^{\barn}$, 
with $\barn = O(\tn \cdot (1/\epsilon)^4)$, is computable in polynomial time, and it has the property that any ball of radius $(1/2) + \epsilon$ has at most
$O((1/\epsilon)^2)$ codewords. Similarly to function $b$, we define the function $c(\barx, r) = $ the $r$-th bit of
${\rm Code}(\barx)$, for $\barx \in \zo^{\ell n}$ and any binary string $r$ with length 
$|r| = \log( {\rm Code}(\barx)) = \log (\ell \cdot N_1 \cdot (1/\epsilon)^4) + O(1)$. We
define the extractor $E'$ by
\begin{equation}
\label{e:localextractor2}
E'(X, (\barx, r)) = c(\barX(\barx), r) \odot  c(\barX(\barx+1), r) \odot \ldots \odot  c(\barX(\barx+m-1), r).
\end{equation} 
The analysis is very similar to that done for the previous extractor given in Equation~(\ref{e:localextractor}). For example, if we assume that
$\epsilon \leq 2^{(1/4)n}$,
take $N_1 = 2^{n^2}$ and $m= 2^{(1/3)n}$, and we denote the length of $X$ by $\barN$ (\ie, $\barN = 2^{n^2 + n}$), we obtain a quite simple extractor that has seed length $O(\log (\barN))$, is capable to extract from sources 
with min-entropy $\lambda \barN$, for arbitrary constant $\lambda > 0$, and has output 
length $\approx 2^{(1/3)\sqrt{\log(\barN)}}$.
This extractor has a good seed length, however the output length is much smaller than the 
min-entropy of the source.

\section{Acknowledgments}
I am grateful to Luca Trevisan for his insightful comments on an earlier draft of this work.



\end{document}